\documentclass[aps,preprintnumbers,superscriptaddress,showpacs,twocolumn,longbibliography]{revtex4}
\usepackage{booktabs}
\usepackage{epsfig}
\usepackage{psfrag}
\usepackage{amsfonts}
\usepackage{amsmath}
\usepackage{graphicx}
\usepackage{float}
\usepackage{subfig}
\usepackage{dcolumn}
\usepackage{bm}

\begin{document}

\title{Study on the $\Omega_{c}^{0}$ States Decaying to $\Xi_{c}^{+}K^{-}$ in $pp$ collisions at
$\sqrt{s}=7$, 13~TeV}

\author{Hong-ge Xu$^{1,2}$, Gang Chen$^{2}$\footnote{Corresponding Author: chengang1@cug.edu.cn},Yu-Liang Yan$^3$, Dai-Mei Zhou$^4$, Liang Zheng$^{2}$, Yi-Long Xie$^{2}$, Zhi-Lei She$^2$ and Ben-Hao Sa$^3$}

\address{
${^1}$Institute of Geophysics and Geomatics, China University of Geosciences, Wuhan 430074, China\\
${^2}$School of Mathematics and Physics, China University of Geoscience, Wuhan 430074, China\\
${^3}$China Institute of Atomic Energy, P.O. Box 275(10), Beijing 102413, China\\
${^4}$Institute of Particle Physics, Huazhong Normal University, Wuhan 430082, China}

\begin{abstract}
The production of strange particles $\Xi_{c}^{+}, K^{-}$ is simulated in mid-rapidity $pp$ collisions at $\sqrt{s}=7$ TeV with $0.2 \leq pt \leq 6$~GeV/c using the {\footnotesize PACIAE} model. The results are consistent with LHCb
experimental data on $\Xi_{c}^{+}$ and $K^{-}$ yield. Then, a dynamically constrained phase-space coalescence ({\footnotesize DCPC}) model plus {\footnotesize PACIAE} model was used to produce the $\Xi_{c}^{+}K^{-}$ bound states and study the narrow excited $\Omega_{c}^{0}$ states through $\Omega_{c}^{0} \to \Xi_{c}^{+}K^{-}$ in $pp$ collisions at $\sqrt{s}=7$ and 13 TeV. The yield, transverse momentum distribution, and rapidity distribution of the five new excited $\Omega_{c}^{0}$ states of $\Omega_{c}(3000)^{0}$, $\Omega_{c}(3050)^{0}$, $\Omega_{c}(3066)^{0}$,$\Omega_{c}(3090)^{0}$and $\Omega_{c}(3119)^{0}$ were predicted.

\end{abstract}

\maketitle

\section{Introduction}
In the early 1960s, many strongly interacting particles were observed in particle$/$nucleon experiments, which were named as "hadron" by L.B. Okun later~\cite{Ref1}. According to these observations, M. Gell-Mann and G. Zweig independently proposed the quark model that is classification scheme for hadrons~\cite{Ref2,Ref3}. Quark model achieves a great success, and it is a milestone in the development of particle physics. A well-known example was that, after the $\Omega$ had been predicted in 1961 independently by M. Gell-Mann~\cite{Ref6} and Y. Ne'eman~\cite{Ref7}, this particle was discovered in 1964~\cite{Ref5}. In the traditional quark model, hadrons can be categorized into two families: baryons made of three quarks and mesons made of one quark and one antiquark. Both mesons and baryons are color singlets. During the last four decades, baryons containing heavy quarks have been the focus of much attention, especially since the development of the efficient theory of heavy quarks and its application to baryons containing a single heavy quark. In recent years, a variety of theories and experiments have been proposed for the study of heavy flavor baryon.
Heavy quark provides a "flavor tag" that can be used as a window into the depths of the color confinement, or at least a window allowing us see further under the nonperturbative QCD epidermis than the light baryons do. In the process of establishing cognition of different energy scale QCD, a rich dynamical study on heavy flavor baryons and their properties is urgently needed.

In the past three decades, various phenomenological models have been used to study heavy baryons, including the relativized potential quark model~\cite{Ref8}, the Feynman-Hellmann theorem~\cite{Ref9}, the combined expansion in $1/m_{Q}$  and $1/N_{c}$~\cite{Ref10}, the relativistic quark model~\cite{Ref11}, the chiral quark model~\cite{Ref12}, the hyperfine interaction~\cite{Ref13,Ref14}, the pion induced reactions \cite{Ref15}, the variational approach~\cite{Ref16}, the Faddeev approach~\cite{Ref17}, the constituent quark model~\cite{Ref18}, the unitarized dynamical model~\cite{Ref19}, the extended local hidden gauge approach~\cite{Ref20}, the unitarized chiral perturbation theory~\cite{Ref21}, etc. There are also many Lattice QCD studies~\cite{Ref22,Ref23}, et al.

Furthermore, there are numerous experimental groups to investigate heavy baryons. Some mass spectra, width, lifetime, decays and form factors of heavy baryons have been reported but the spin and parity identification of some states are still missing. By now, all the ground state charmed baryons containing a single charm quark have been well established both experimentally and theoretically~\cite{Ref24}. The lowest-lying orbitally excited charmed states $\Lambda_{c}(2959)^{0}(J^{P}=1/2^{-})$, $\Lambda_{c}(2625)^{0}(J^{P}=3/2^{-})$, $\Xi_{c}(2790)^{0}(J^{P}=1 /2^{-})$ and $\Xi_{c}(2815)^{0}(J^{P}=3/2^{-})$ have been well observed by several collaborations, which made the two $SU(4)$ multipletes complete~\cite{Ref25,Ref26,Ref27,Ref28,Ref29}. Besides that, several P-wave charm baryon candidates $\sum_{c}(2800)$, $\Xi_{c}(2980)$ and $\Xi_{c}(3080)$ were also well observed by the Belle and BABAR collaborations \cite{Ref29,Ref30,Ref31,Ref32}. In 2017, the LHCb collaboration reported their observation of five new narrow excited $\Omega_{c}^{0}$ states decaying to $\Xi_{c}^{+}K^{-}$, based on samples of $pp$ collision data corresponding to integrated luminosities of 1.0, 2.0, and 0.3 fb$^{-1}$ at center-of-mass energies of 7, 8, and 13 TeV, respectively~\cite{Ref33}. In future, the experiments at JPARC, PANDA~\cite{Ref34} and LHCb are expected to give further information on charmed baryons pretty soon.

These heavy baryons provide us an ideal platform to deepen our understanding of the non-perturbative QCD. Therefore, people hope to fully understand their nature. In this paper, we study the $\Xi_{c}^{+}K^{-}$ bound state to predict the properties of $\Omega_{c}^{0}$ states decaying to $\Xi_{c}^{+}K^{-}$ by simulating analysis. First, we generate $pp$ collision events at $\sqrt s=7$ and 13~TeV to obtain the hadrons of $\Xi_{c}^{+}$ and $K^{-}$ using the parton and hadron cascade model ({\footnotesize PACIAE})~\cite{Ref35}. Then we use a dynamically constrained phase space coalescence model ({\footnotesize DCPC})~\cite{Ref36,Ref37,Ref38} to produce $\Xi_{c}^{+}K^{-}$ bound states for the study of $\Omega_{c}^{0}$. Here, we mainly simulate and study five different excited resonance $\Omega_{c}^{0}$ states of $\Omega_{c}(3000)^{0}$, $\Omega_{c}(3050)^{0}$, $\Omega_{c}(3066)^{0}$, $\Omega_{c}(3090)^{0}$ and $\Omega_{c}(3119)^{0}$, which is observed by LHCb experiment.

\section{{\footnotesize PACIAE} Model and {\footnotesize DCPC} Model}
The parton and hadron cascade model {\footnotesize PACIAE}~\cite{Ref35} is based on {\footnotesize PYTHIA} 6.4 to simulate various collision, such as $e^{+}e^{-}$, $pp$, $p$-$A$ and $A$-$A$ collisions. In general, {\footnotesize PACIAE} has four main physics stages consisting of the parton initiation, parton rescattering, hadronization, and hadron rescattering. In the parton initiation, the string fragmentation is switched off temporarily in {\footnotesize PACIAE} and di(anti)quarks are broken into (anti)quarks. This partonic initial state can be regarded as quark-gluon matter (QGM) formed inside the parton initialization stage in the $pp$ collisions. Then the parton rescattering in QGM is taken into account by the $2\to 2$ LO-pQCD parton-parton cross sections\cite{PYTHIA6.4}. Their total and differential cross sections in the parton evolution is computed by the Monte Carlo method. In the hadronization process, the parton can be hadronized by the Lund string fragmentation regime and/or the phenomenological coalescence model~\cite{PYTHIA6.4}. The final stage is the hadron rescattering process happening between the created hadrons until the hadronic freeze-out.

In the theoretical papers, the yield of nuclei (bound states) usually is calculated in two steps: First, the nucleons are calculated by the transport model. Then, the nuclei are calculated by the phase-space coalescence model based on the Wigner function~\cite{New45,New452} or by the statistical model~\cite{New46}. We proposed a dynamically constrained phase-space coalescence ({\footnotesize DCPC}) model to calculate the yield of nuclei(bound states) after the transport model simulations.

According to the {\footnotesize DCPC} model, the yield of a single particle is estimated by
\begin{equation}
Y_1=\int_{H\leq E} \frac{d\vec qd\vec p}{3h},
\end{equation}
where $E$ and $H$ denote energy and the Hamiltonian of the particle, respectively. Furthermore, the yield of a cluster consisting of $N$ particles is defined as following:
\begin{equation}
Y_{N} =\int\cdots\int_{H\leq{E}}{\frac{d\vec q_{1}d\vec p_{1}\cdots d\vec q_{N}d\vec p_{N}}{(3h)^{N}}}.
\end{equation}

Therefore, the yield of a $\Xi_{c}^{+}K^{-}$ cluster in the {\footnotesize DCPC} model can be calculated by
\begin{align}
Y_{\Omega_{c}^{0}} =&\int ...\int\delta_{12}\frac{d\vec q_1d\vec p_1 d\vec q_2d\vec p_2}{(3h)^{2}},
\label{yield} \\
\delta_{12}=&\left\{
  \begin{array}{ll}
  1 \hspace{0.2cm} \textrm{if} \hspace{0.2cm} 1\equiv \Xi_{c}^{+}, 2\equiv K^{-};\\
    \hspace{0.2cm} m_{\Omega_{c}^{0}}-\Delta m\leq m_{inv}\leq m_{\Omega_{c}^{0}}+\Delta m,
q_{12}\leq D_{0};\\
  0 \hspace{0.2cm}\textrm{otherwise}.
  \end{array}
  \right.
\label{yield1}
\end{align}
\textrm{where},
\begin{equation}
\hspace{0.5cm}  m_{inv}=\sqrt{(E_1+E_2)^2-(\vec p_1+\vec p_2)^2}.
\label{yield2}
\end{equation}
 The $q_{12}$ is the distance between the two particles ($\Xi_{c}^{+}$ and $K^{-}$ ), $m_{\Omega_{c}^{0}}$ denotes the mass of $\Omega_{c}^{0}$, and $\Delta m$ refers to its mass uncertainty. $E_1, E_2$ and $p_{1}, p_2$ denote the energies and momenta of the two particles ($\Xi_{c}^{+}$ and $K^{-}$). Here, we assume that the diameter of the bound state $\Xi_{c}^{+}K^{-}$ is $D_{0}=1.74$ fm.

\section{Results}
We first simulate the $pp$ collision events using the {\footnotesize PACIAE} model at $\sqrt{s}= 7$ and 13~TeV. The capability of {\footnotesize PACIAE} to describe the generation of the final state particles in $pp$ collisions has been detailed in Refs.~\cite{Ref36,Ref37,Ref38,Ref41,Ref42,Ref43,Ref44}. In order to obtain a suitable set of model parameters, the results on the yield of $\Xi_{c}^{+}$ and $K^{-}$ were roughly fitted to the LHCb data in $pp$ collisions at $\sqrt{s}= 7$~TeV\cite{Ref47,Ref48,Ref49,Ref50,Ref51}. The simulation results with {\footnotesize PACIAE} agree well with the experimental results, as shown in Table I. It should be said that the yield of $\Xi_{c}^{+}$ for the experiment data in the Table I is calculated by the ratio of the cross sections of $\Xi_{c}^{0}$ to the $D_{0}$ meson and the ratio of $\Xi_{c}^{0}$ to $\Xi_{c}^{+}$ according to the data in Ref.\cite{Ref48,Ref49,Ref50,Ref51}.

\begin{large}
\small\addtolength{\tabcolsep}{12.pt}
\begin{table}[hp]
\caption{~The yield of $\Xi_{c}^{+}$ and $K^{-}$ computed by {\footnotesize PACIAE} in mid-rapidity $pp$ collisions at $\sqrt{s}= 7$~TeV with $0.2 \leq p_t \leq 6 $~GeV/c and comparison with experiment data~\cite{Ref47,Ref48,Ref49,Ref50,Ref51}.}
\begin{tabular}{cccc} \hline  \hline
particles &{\footnotesize PACIAE}  & Experiment data  \\ \hline
$K^{-}$   & 0.286  & $0.286\pm0.016$\\
$\Xi_{c}^{+}$ &$7.40\times 10^{-5}$&$7.47\pm0.14\times 10^{-5}$\\ \hline  \hline
\end{tabular} \label{paci1}
\end{table}
\end{large}

In this work, we assume that the narrow excited $\Omega_{c}^{0}$ states ($\Omega_{c}(3000)^{0}$, $\Omega_{c}(3050)^{0}$, $\Omega_{c}(3066)^{0}$, $\Omega_{c}(3090)^{0}$ and $\Omega_{c}(3119)^{0}$) are the $\Xi_{c}^{+}K^{-}$ bound state generated through $\Omega_{c}^{0} \to \Xi_{c}^{+}K^{-}$, which is produced during the hadron evolution period. We use {\footnotesize PACIAE} transport model to generate 300 million events of $pp$ collision at $\sqrt {s}=7$ and 13 TeV, and input the final state particles $\Xi_{c}^{+}$ and $K^{-}$ into {\footnotesize DCPC} model to construct the clusters of $\Xi_{c}^{+}K^{-}$, as the Eq.(3) and (4). Then the characteristics of the narrow excited $\Omega_{c}^{0}$ states of $\Omega_{c}(3000)^{0}$, $\Omega_{c}(3050)^{0}$, $\Omega_{c}(3066)^{0}$, $\Omega_{c}(3090)^{0}$ and $\Omega_{c}(3119)^{0}$ can be studied.

\begin{figure*}[t]
\includegraphics[width=0.4\textwidth]{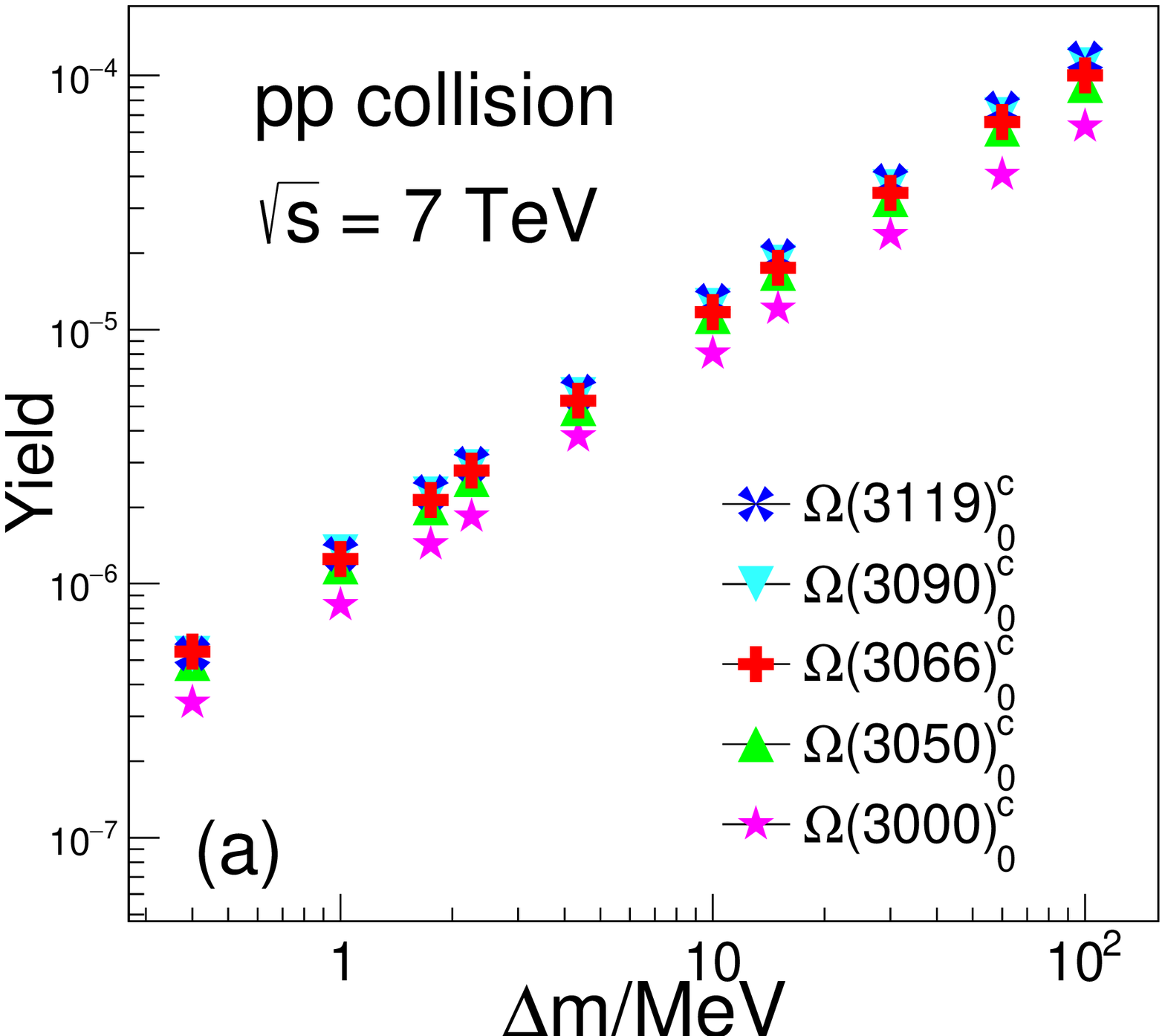}
\includegraphics[width=0.4\textwidth]{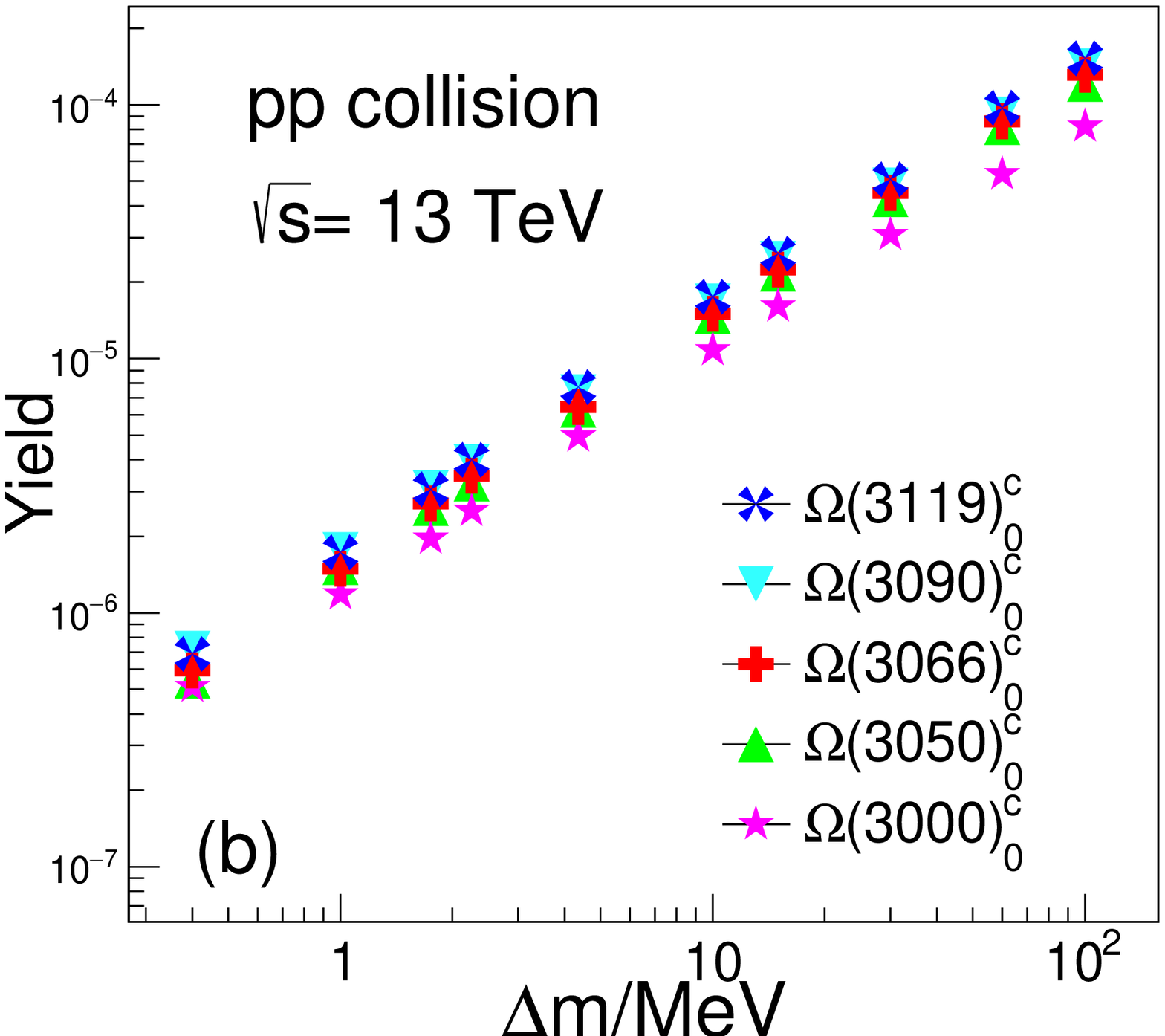}
\caption{Logarithmic distribution of the yields of the five narrow excited resonant $\Omega_{c}^{0}$ states, as a function of $\Delta m$, in $pp$ collisions, (a) at $\sqrt {s}=7$~TeV, (b) at $\sqrt {s}=13$~TeV. The data are calculated using {\footnotesize PACIE+DCPC} model as $\Omega_{c}^{0}$ states decaying to $\Xi_{c}^{+}K^{-}$ bound states.}
\label{tu1}
\end{figure*}

\begin{large}
\small\addtolength{\tabcolsep}{1.pt}
\begin{table*}[tbp]
\caption{The yields of five new resonant $\Omega_{c}^{0}$ states varies with $\Delta m$ from 0.4 MeV to 100 MeV in $pp$ collision at $\sqrt{s}= 7$ and 13~TeV, the date are obtained by $\Omega_{c}^{0}$ states decaying to $\Xi_{c}^{+}K^{-}$ bound states which compute using {\footnotesize PACIE+DCPC} model.}
\begin{tabular}{ccccccccccc} \hline\hline
$\Delta m$ &\multicolumn{5}{c}{$\sqrt{s}= 7$~TeV} &\multicolumn{5}{c}{$\sqrt{s}= 13$~TeV}\\
\cmidrule[0.0pt](l{0.01cm}r{0.01cm}){2-6}
\cmidrule[0.0pt](l{0.01cm}r{0.01cm}){7-11}
  (MeV)&$\Omega_{c}(3000)^{0}$&$\Omega_{c}(3050)^{0}$ &$\Omega_{c}(3066)^{0}$ &$\Omega_{c}(3090)^{0}$ &$\Omega_{c}(3119)^{0}$ &$\Omega_{c}(3000)^{0}$ &$\Omega_{c}(3050)^{0}$ &$\Omega_{c}(3066)^{0}$ &$\Omega_{c}(3090)^{0}$ &$\Omega_{c}(3119)^{0}$\\
\hline

0.4 &3.40$\times10^{-7}$ &4.87$\times10^{-7}$ &5.4$\times10^{-7}$ &5.30$\times10^{-7}$ &5.33$\times10^{-7}$ &5.1$\times10^{-7}$&5.40$\times10^{-7}$&5.93$\times10^{-7}$& 7.20$\times10^{-7}$ &6.90$\times10^{-7}$\\

1& 8.23$\times10^{-7}$ &  1.17$\times10^{-6}$& 1.25$\times10^{-6}$& 1.33$\times10^{-6}$ &1.31$\times10^{-6}$& 1.18$\times10^{-6}$& 1.51$\times10^{-6}$&  1.49$\times10^{-6}$&  1.77$\times10^{-6}$ &1.73$\times10^{-6}$ \\

 1.75& 1.43$\times10^{-6}$ & 2.00$\times10^{-6}$&2.14$\times10^{-6}$&2.25$\times10^{-6}$&2.30$\times10^{-6}$ & 1.96$\times10^{-6}$&2.59$\times10^{-6}$& 2.69$\times10^{-6}$& 3.10$\times10^{-6}$ &3.07$\times10^{-6}$ \\

  2.25& 1.84$\times10^{-6}$ & 2.58$\times10^{-6}$&2.79$\times10^{-6}$&2.88$\times10^{-6}$ &2.96$\times10^{-6}$ & 2.52$\times10^{-6}$&3.22$\times10^{-6}$& 3.47$\times10^{-6}$& 3.93$\times10^{-6}$  &4.00$\times10^{-6}$\\

  4.35 & 3.79$\times10^{-6}$ & 4.91$\times10^{-6}$& 5.26$\times10^{-6}$&5.53$\times10^{-6}$ &5.71$\times10^{-6}$ & 4.94$\times10^{-6}$&6.30$\times10^{-6}$&6.46$\times10^{-6}$& 7.40$\times10^{-6}$ &7.74$\times10^{-6}$ \\

  10 & 8.04$\times10^{-6}$ & 1.13$\times10^{-5}$& 1.17$\times10^{-5}$&1.24$\times10^{-5}$ &1.30$\times10^{-5}$& 1.08$\times10^{-5}$&1.47$\times10^{-5}$& 1.50$\times10^{-5}$& 1.68$\times10^{-5}$ &1.75$\times10^{-5}$\\

  15 & 1.21$\times10^{-5}$ & 1.67$\times10^{-5}$& 1.75$\times10^{-5}$&1.83$\times10^{-5}$ &1.95$\times10^{-5}$& 1.61$\times10^{-5}$&2.17$\times10^{-5}$& 2.25$\times10^{-5}$& 2.47$\times10^{-5}$ &2.59$\times10^{-5}$\\

  30  & 2.36$\times10^{-5}$ & 3.27$\times10^{-5}$& 3.46$\times10^{-5}$&3.64$\times10^{-5}$ &3.84$\times10^{-5}$& 3.08$\times10^{-5}$&4.26$\times10^{-5}$& 4.49$\times10^{-5}$& 4.80$\times10^{-5}$ &5.08$\times10^{-5}$ \\

  60 & 4.08$\times10^{-5}$ & 6.19$\times10^{-5}$& 6.56$\times10^{-5}$&6.99$\times10^{-5}$ &7.44$\times10^{-5}$ & 5.32$\times10^{-5}$&8.16$\times10^{-5}$& 8.59$\times10^{-5}$& 9.15$\times10^{-5}$ &9.72$\times10^{-5}$\\

  100 & 6.32$\times10^{-5}$ & 9.19$\times10^{-5}$& 1.00$\times10^{-4}$&1.10$\times10^{-4}$&1.17$\times10^{-4}$ & 8.22$\times10^{-5}$&1.21$\times10^{-4}$& 1.31$\times10^{-4}$& 1.42$\times10^{-4}$ &1.52$\times10^{-4}$\\ \hline\hline
\end{tabular} \label{paci1}
\end{table*}\end{large}

Fig.1 shows the yield distributions of excited resonance states $\Omega_{c}^{0}$ in full rapidity phase space with $\Delta m$ varying from 0.4 MeV to 100 MeV in $pp$ collisions at $\sqrt {s}=$7 TeV and 13 TeV,
including resonance states $\Omega_{c}(3000)^{0}$, $\Omega_{c}(3050)^{0}$, $\Omega_{c}(3066)^{0}$, $\Omega_{c}(3090)^{0}$ and $\Omega_{c}(3119)^{0}$, respectively. The values are shown in Table II.

From the Fig.1, we can see that the yield $\ln{Y}$ of the four excited resonant $\Omega_{c}^{0}$ states computed using {\footnotesize PACIAE+DCPC} model increase linearly with parameter ln${\Delta m}$, while $\Delta m$ changes from 0.4~MeV to 100~MeV. The values are on the order of $10^{-7}$ to $10^{-4}$. The yield of $\Omega_{c}^{0}$ in $pp$ collision of $\sqrt{s}= 13$~TeV is more than the yield of it at $\sqrt{s}= 7$~TeV.

In the LHCb experiment, the five new, narrow excited $\Omega_{c}^{0}$ states of the $\Omega_{c}(3000)^{0}$, $\Omega_{c}(3050)^{0}$, $\Omega_{c}(3066)^{0}$, $\Omega_{c}(3090)^{0}$, and $\Omega_{c}(3119)^{0}$ are observed and measurements of their masses and decaying widths are given separately~\cite{Ref33}. If we take half of the decay width of their mass as $\Delta m$ parameter, then we may predict their yields, as shown in Table III.

\begin{large}
\small\addtolength{\tabcolsep}{3.pt}
\begin{table}[htp]
\caption{ The value of the five new excited resonant $\Omega_{c}^{0}$ states in $pp$ collision at $\sqrt{s}= 7$ and 13~TeV, computed using {\footnotesize PACIAE+DCPC} model, where $\Delta m = \Gamma/2$~\cite{Ref33}.}
\begin{tabular}{cccc} \hline  \hline
Resonance &$\Delta m$(MeV)&Yield(7~TeV)& Yield(13~TeV) \\ \hline
$\Omega_{c}(3000)^{0}$&$2.25\pm0.30$ & $1.84\times10^{-6}$&$2.52\times10^{-6}$\\
$\Omega_{c}(3050)^{0}$&$0.40\pm0.10$ &$4.87\times10^{-7}$ &$5.40\times10^{-7}$\\
$\Omega_{c}(3066)^{0}$&$1.75\pm0.20$ &$2.14\times10^{-6}$ &$2.69\times10^{-6}$\\
$\Omega_{c}(3090)^{0}$&$4.35\pm0.52$ & $5.53\times10^{-6}$&$7.40\times10^{-6}$\\
$\Omega_{c}(3119)^{0}$&$0.55\pm0.40$ & $7.13\times10^{-7}$&$9.57\times10^{-7}$\\
\hline  \hline
\end{tabular} \label{paci1}
\end{table}\end{large}

Fig.2 shows the transverse momentum $p_{T}$ distributions of the five excited resonant $\Omega_{c}^{0}$ states of $\Omega_{c}(3000)^{0}$, $\Omega_{c}(3050)^{0}$, $\Omega_{c}(3066)^{0}$, $\Omega_{c}(3090)^{0}$) and $\Omega_{c}(3119)^{0}$ in $pp$ collision at $\sqrt{s}= 7$ and 13~TeV. In each panel, the red dashed line refers to the distribution in $pp$ collision at $\sqrt{s}=$ 7~TeV, the blue solid line corresponding to the $\sqrt{s}=13$~TeV. The peak of $p_{T}$ distributions is about 1.5 GeV. It can be seen from this figure that all the transverse momentum distribution characteristics of the produced excited resonant $\Omega_{c}^{0}$ states are similiar. In the same way, the transverse momentum distribution characteristics of the produced excited resonant $\Omega_{c}^{0}$ states is not sensitive to the reaction energy of $pp$ collision.
\begin{figure*}[tb]
\centering
\includegraphics[width=0.8\textwidth]{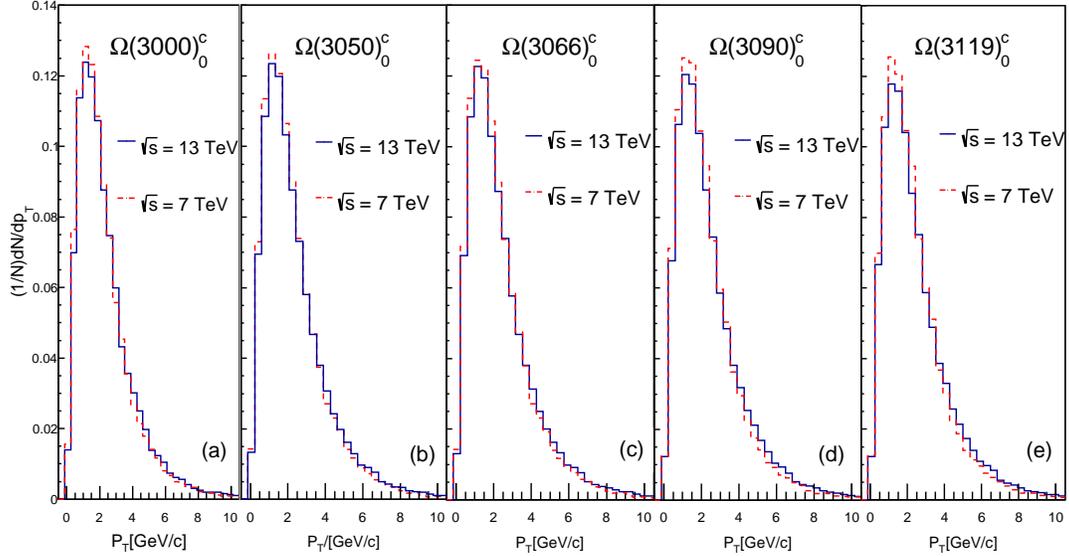}
\caption{(Color online) Transverse momentum distributions of the five excited resonant $\Omega_{c}^{0}$ states by the decay from $\Omega_{c}^{0} \to \Xi_{c}^{+}K^{-}$ in $p p$ collision at $\sqrt{s}= 7$ (red dashed histograms) and 13~TeV (blue solid histograms) calculated by the final hadronic states in the {\footnotesize {PACIAE+DCPC}} model simulations with $\Delta m= 0.1$ GeV, respectively.}
\label{tu3}
\end{figure*}
\begin{figure*}[!htb]
\includegraphics[width=0.8\textwidth]{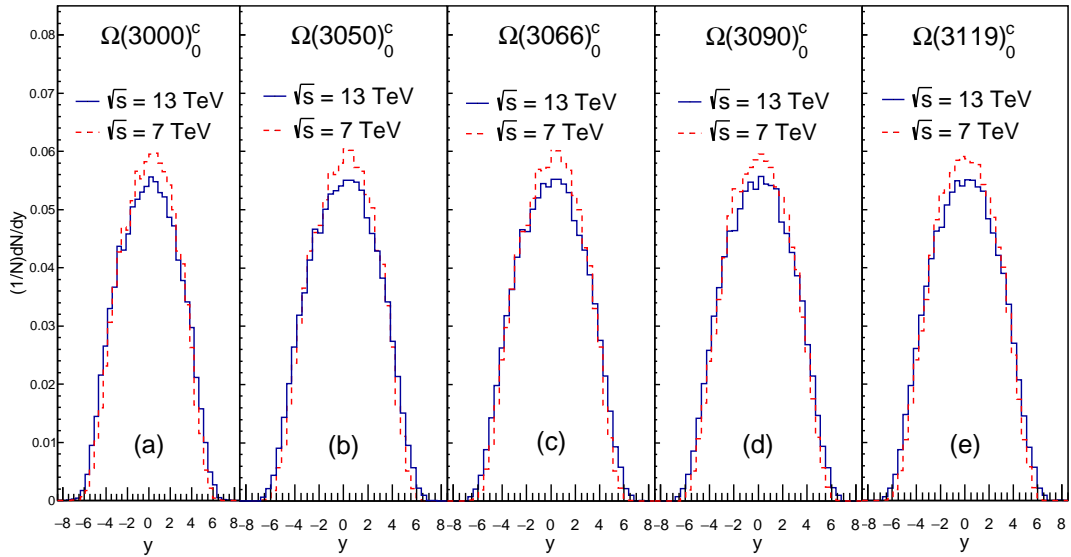}
\centering
\caption{(Color online) Similar to Fig. 2, but for the rapidity distribution.}
\label{tu4}
\end{figure*}
 We also predicted the rapidity distribution of excited resonant $\Omega_{c}^{0}$ states ($\Omega_{c}^{0}$ states of $\Omega_{c}(3000)^{0}$, $\Omega_{c}(3050)^{0}$, $\Omega_{c}(3066)^{0}$, $\Omega_{c}(3090)^{0}$ and $\Omega_{c}(3119)^{0}$ in $pp$ collision at $\sqrt{s}=7$ and 13~TeV shown in Fig3. From this figure, one sees that the global features of rapidity distributions are similar to the different excited resonant $\Omega_{c}^{0}$ states particles and different collision energies. But the rapidity distribution of $\sqrt{s}= 13$ TeV is slightly wider than the distribution at $\sqrt{s}= 7$ TeV. The rapidity distribution is symmetry and the range is about from 6.5 to -6.5.

\section{conclusions}
In this paper we simulate the generation of final state particles in $pp$ collisions at $\sqrt{s}= 7$ and 13~TeV with {\footnotesize PACIAE} Model, and study the production of strange particles $\Xi_{c}^{+}$ and $K^{-}$, which are consistent with the data of LHCb. Then the $\Xi_{c}^{+}$ and $K^{-}$ are input into the {\footnotesize DCPC} model to construct the clusters of $\Xi_{c}^{+}K^{-}$, and the characteristics of the new narrow excited $\Omega_{c}(3000)^{0}$, $\Omega_{c}(3050)^{0}$, $\Omega_{c}(3066)^{0}$, $\Omega_{c}(3090)^{0}$ and $\Omega_{c}(3119)^{0}$ are studied, based on the assumption that the new narrow excited $\Omega_{c}^{0}$ states is $\Xi_{c}^{+}K^{-}$ bound state. The results show that the transverse momentum distribution characteristics of the five different excited resonant $\Omega_{c}^{0}$ states of $\Omega_{c}(3000)^{0}$, $\Omega_{c}(3050)^{0}$, $\Omega_{c}(3066)^{0}$, $\Omega_{c}(3090)^{0}$ and $\Omega_{c}(3119)^{0}$ are all the similar, and so is the rapidity. If we take half of the width of the mass decay as a parameter $\Delta m$, we may predict that the yield of the new narrow excited resonant $\Omega_{c}(3000)^{0}$, $\Omega_{c}(3050)^{0}$, $\Omega_{c}(3066)^{0}$, $\Omega_{c}(3090)^{0}$) and $\Omega_{c}(3119)^{0}$ computed in $pp$ collisions at $\sqrt{s}= 7$ is $1.84\times10^{-6}$, $0.487\times10^{-6}$, $2.14\times10^{-6}$, $5.53\times10^{-6}$, and $0.71\times10^{-6}$, respectively. As well as, when the reaction energy is 13 TeV, their corresponding yield is $2.52\times10^{-6}$, $0.54\times10^{-6}$, $2.69\times10^{-6}$, $7.40\times10^{-6}$, and $0.96\times10^{-6}$, respectively. Obviously, the yield of these particles increases as the reaction energy increases.

The study of the new narrow excited $\Omega_{c}^{0}$ states productions in $pp$ collisions is under way. One may expect more diverse results in $pp$ or Nucleus-nucleus collisions.

\section {ACKNOWLEDGMENT}
The authors thank Prof. Larisa V. Bravina (University of Oslo) for valuable comments. This work is supported by the NSFC (11475149, 11705167).

\end{document}